\documentclass[preprintnumbers,nofootinbib,superscriptaddress,10pt]{revtex4}
\usepackage[dvipdfmx]{graphicx} 
\usepackage{amsmath,amssymb,amsfonts, bm} 

\def\a{\alpha} \def\b{\beta}   \def\d{\delta} \def\D{\Delta} \def\e{\epsilon}  \def\h{\eta} \def\th{\theta}     \def\m{\mu} \def\n{\nu}     \def\r{\rho} \def\s{\sigma}  \def\t{\tau}     \def\c{\chi}  

\def\dg{\dagger}  \def\nn{\nonumber}

\newcommand{\lsp}{ \left ( } \newcommand{\rsp}{ \right ) }   \newcommand{\To}{\Rightarrow}

 \renewcommand{\Im}{{\rm Im \, }}

\newcommand{\diag}[2]{ \begin{pmatrix}  #1 & 0 \\ 0 & #2 \\   \end{pmatrix}  }

\newcommand{\Diag}[3]{ \begin{pmatrix} #1 & 0 & 0 \\ 0 & #2 & 0 \\ 0 & 0 & #3 \\\end{pmatrix}}

\usepackage{color}

\begin{document}

\title{\large 
Almost general analysis of $\mu-\tau$ reflection symmetry perturbed by charged leptons and its testability by DUNE and T2HK
}

\preprint{STUPP-25-277}

\author{Masaki J. S. Yang}
\email{mjsyang@mail.saitama-u.ac.jp}
\affiliation{Department of Physics, Saitama University, 
Shimo-okubo, Sakura-ku, Saitama, 338-8570, Japan}
\affiliation{Department of Physics, Graduate School of Engineering Science,
Yokohama National University, Yokohama, 240-8501, Japan}



\begin{abstract} 

In this paper, we generally analyze the $\mu - \tau$ reflection symmetry modified by small mixings of  charged leptons and 
how will future experiments verify deviations from the predictions of the symmetry. 
As an approximation, the left-handed diagonalization $U_{e}$ of charged leptons
 is assumed to have a similar magnitude as the CKM matrix. In other words,  the 1-3 mixing is neglected and the 1-2 and 2-3 mixing are to be approximately $O(0.1)$.  The Dirac CP phase $\delta$ of the MNS matrix is evaluated in such parameter regions. 
As a result, deviations from the predictions $\sin \theta_{23} = \pi/4$ and $\delta = \pm \pi/2$ depend on  relative CP phases between $U_{e}$ and diagonalization of neutrinos $U_{\nu}$.
While phases of the second and third generations cause only about $\pm 10^{\circ}$ deviations for the Dirac phase $\delta$, the phase of the first generation can cause up to $\pm 30^{\circ}$.
This flavor dependence is distinguished to some extent by the next-generation experiments.
On the other hand, if $\delta$ is not observed, such a scenario is excluded by about 5 years of observation.

\end{abstract} 

\maketitle

\section{Introduction}

Understanding of CP violation (CPV) in the lepton sector is an crucial issue in studies of flavor physics and leptogenesis \cite{Fukugita:1986hr}.
Although observations of the Dirac phase $\d$ have been continued by the T2K \cite{T2K:2021xwb} and the NO$\nu$A experiment \cite{NOvA:2021nfi}, the value is rarely determined in the global fit of the normal mass hierarchy \cite{Esteban:2024eli}.
As next-generation experiments, Deep Underground Neutrino Experiment (DUNE) \cite{DUNE:2020jqi} and Tokai to Hyper-Kamiokande (T2HK) \cite{Hyper-KamiokandeProto-:2015xww} are scheduled in the next decade. 
These experiments can discover up to $|\sin \d| \gtrsim 0.5$ after about 10 years of observation, with a resolution of $\pm 20^{\circ}$ around the maximum value $\pm 90^{\circ}$ \cite{Walker:2019cxi, DUNE:2020jqi}.

Meanwhile, to explain certain Dirac and Majorana phases, there have been a number of literatures of  generalized CP symmetry (GCP) \cite{Ecker:1981wv, Ecker:1983hz, Gronau:1985sp, Ecker:1987qp,Neufeld:1987wa,Ferreira:2009wh,Feruglio:2012cw,Holthausen:2012dk,Ding:2013bpa,Girardi:2013sza,Nishi:2013jqa,Ding:2013hpa,Feruglio:2013hia,Chen:2014wxa,Ding:2014ora,Ding:2014hva,Chen:2014tpa,Li:2015jxa,Turner:2015uta, Rodejohann:2017lre, Penedo:2017vtf, Nath:2018fvw,  Yang:2021smh, Yang:2021xob, Ohki:2023zsn}. 
A notable example is the $\m-\t$ reflection symmetry 
\cite{Harrison:2002et,Grimus:2003yn, Grimus:2005jk, Joshipura:2007sf, Adhikary:2009kz, Joshipura:2009tg, Xing:2010ez, Ge:2010js, He:2011kn, Gupta:2011ct, Grimus:2012hu, He:2012yt, Joshipura:2015dsa, Xing:2015fdg, He:2015afa, Chen:2015siy, He:2015xha, Samanta:2017kce,  Nishi:2018vlz, Sinha:2018xof, Huang:2018fog, Xing:2019edp, Pan:2019qcc, Chakraborty:2019rjc, Liao:2019qbb, Yang:2020qsa, Zhao:2021dwc, Bao:2022kon, Xing:2022uax, Zhao:2024zgx, Huang:2024xiq, Kumar:2024zfb}
that predicts the maximal 2-3 mixing $\th_{23} = \pi /4$, the maximal Dirac phase $\d = \pm \pi/2$ and Majolana phases 0 or $\pi/2$ \cite{Xing:2017cwb, Nath:2018hjx}. 
The breaking of the $\m-\t$ reflection symmetry and its observability are also widely discussed
\cite{Zhao:2017yvw, Liu:2017frs, Nath:2018xkz, Duarah:2020zjo, Hyodo:2025xug}. 

When the Dirac phase is observed in the near future,
it is generally expected to deviate from the maximal value $\d = \pm \pi/2$. 
In this paper, we generally analyze how the prediction of $\m-\t$ reflection symmetry is modified by small mixings  of charged leptons and  is verified in the current experimental circumstances. 

This paper is organized as follows. 
The next section gives a review of $\m - \t$ reflection symmetry and representation of the MNS matrix. 
In Sec.~3, we analyze the future verifiability of this setup.
The final section is devoted to a summary. 

\section{$\mu - \tau$ reflection symmetry and perturbation by charged leptons}

In this section, we first review $\mu - \tau$ reflection symmetry \cite{Harrison:2002et,Grimus:2003yn} and organize the notation of  perturbations by charged leptons. 
The $\mu - \tau$ reflection symmetry for the mass matrix of light majorana neutrinos $m_{\n}$ is defined as
\begin{align}
  T m_{\n}^{*} T = m_{\n} \, , ~~~ 
  T = 
\begin{pmatrix}
1 & 0 & 0 \\
0 & 0 & 1 \\
0 & 1 & 0
\end{pmatrix} \, . 
\end{align}
This symmetry restricts $m_{\n}$ to the following form
\begin{align}
m_{\n} = 
\begin{pmatrix}
a & b + i c & b - i c\\
b + i c & d + i f & g  \\
b - i c & g  & d - i f
\end{pmatrix} \, ,
\end{align}
where $a \sim h$ are real parameters.
If the 1-3 mixing of the unitary matrix $U_{\n}$ diagonalizing $m_{\n}$ is not zero, 
it predicts the maximal 2-3 mixing and CP violation 
\begin{align}
\th_{13} \neq 0 ~~ \To ~~ \th_{23} = \pi/4 ~~ {\rm and } ~~ \d = \pm \pi/2 \, . 
\label{eq3}
\end{align}
This symmetry is usually defined in the diagonal basis for the mass matrix of charged leptons $m_{e}$. However, off-diagonal elements are generally introduced in some model-building and
 the left-handed unitary matrix $U_{e}$ in the diagonalization of $m_{e}$ can change these predictions.
Many studies consider perturbations by charged leptons for various characteristic mixing matrices 
\cite{Xing:2005ur, Farzan:2006vj, Hochmuth:2007wq, Ge:2011qn, Ge:2011ih, Dasgupta:2014ula, Petcov:2014laa, Girardi:2014faa, Girardi:2015vha, Delgadillo:2018tza}
to analyze the Dirac phase from this viewpoint. 

Thus, in this paper, we will consider the $\m-\t$ reflection symmetry  defined in a basis where $m_{e}$ is hierarchical and has only small mixing\footnote{
The $\m - \t$ (or 2-3) symmetry in a basis where $m_{e}$ is not diagonal is discussed in Refs.~\cite{Koide:2002cj,Joshipura:2005vy,Yang:2020goc}. Here, Eq.~(\ref{eq4}) is referred to as $\m - \t$ reflection symmetry in a hierarchical basis. }, 
\begin{align}
 T m_{\n}^{*} T = m_{\n} \, , ~~ m_{e} \sim 
\begin{pmatrix}
m_{e11} &  m_{e12}  & m_{e13}  \\
m_{e21} & m_{e22} & m_{e23}  \\
m_{e31} & m_{e32} & m_{e33} \\
\end{pmatrix} ,
\label{eq4}
\end{align}
where matrix elements $m_{eij}$ 
satisfy the following conditions 
\begin{align}
|m_{e 33}| \gg |m_{e23}| ,|m_{e32}|, |m_{e13}| , |m_{e31}| , ~~ {\rm and} ~~ |m_{e22} | \gg |m_{e12}| , |m_{e21}| \, . 
\label{eq5}
\end{align}

In general, unitary matrices for diagonalization $U_{\n , e}$  are represented in the form $U_{\n , e} = \Phi_{L}^{\n , e} U_{\n, e}^{0} \Phi_{R}^{\n, e}$ with diagonal phase matrices $\Phi_{L,R}^{\n, e}$ and 
their PDG parameterization $U_{\n, e}^{0}$
that has three mixing angles and a phase. 
Under these definitions, the MNS matrix is written by
\begin{align}
U_{\rm MNS} & = U_{e}^{\dg} U_{\n}  = \Phi_{R}^{e \dg} U_{e}^{0 \dg} \Phi_{L}^{e \dg} \Phi_{L}^{\n} U_{\n}^{0} \Phi_{R}^{\n} \, .
\end{align}
From Eq.~(\ref{eq3}),  $U_{\n}^{0}$ has the maximal 2-3 mixing and CP phase.
Phase matrices $\Phi_{R}^{\n , e}$ are omitted from now on because they have no physical effect on the Dirac phase $\d$. 

\subsection{Approximation and parameterization for MNS matrix}

Here, we introduce an approximation to the MNS matrix and a specific parameterization to prepare the numerical analysis.

\begin{description}
\item[\bf Approximation:] We neglect the 1-3 mixing in $U_{e}^{0}$, and set the 1-2 mixing $s_{e}$ and 2-3 mixing $s_{\t}$ to about 0.1 or less, $|s_{e,\t}| \lesssim 0.1$.

\item[\bf Justification:] 

The approximate chiral symmetry associated with the first generation suppresses the 1-2 and 1-3 mixings. 
In the limit of zero matrix elements $m_{e 1 i}$ and $m_{e j 1}$, the lightest mass singular value becomes zero and the following $U(1)_{L,R}$ chiral symmetry is restored; 
\begin{align}
 \Diag{e^{- i\a_{L}} }{1}{1} m_{e}  \Diag{e^{i\a_{R}} }{1}{1} = m_{e} \, , 
\end{align}
where $\a_{L,R}$ are real phases. 
A similar symmetry emerges for the second generation. 
Although these symmetries are approximate in reality, mixing angles are suppressed by powers of the ratio $m_{ei}/m_{ej}$ of the corresponding singular values $m_{ei}$. 
For example, when the (1,3) matrix element $m_{e13}$ is of the same order as $m_{e1}$, 
the 1-3 mixing in $U_{e}^{0}$ is roughly evaluated as $m_{e13} / m_{e33} \sim m_{e1} / m_{e3} \simeq 3 \times 10^{-4}$. Therefore, it is safely neglected. 

In the presence of the above chiral symmetries, the CKM matrix $U_{\rm CKM} = U_u^\dagger U_d$ 
can be dominated by down-type quarks, as the corresponding ratios are smaller. 
Since the $d$-$e$ unified relation $m_{e i} \sim m_{d i}$ is well known, 
naive unified theories with the chiral symmetries predict a relation $U_e \sim U_d \sim V_{\rm CKM}$. 

The neglect of 1-3 mixing is also quantitatively justified. In a subsequent work \cite{Yang:2025hex}, a general perturbative analysis including the 1-3 mixing $s_{13}^{e}$ of the charged leptons 
yields the following first-order relation for the Dirac phase; 
\begin{align}
\d & \simeq \delta _{\nu } 
+ s^e_{23} \frac{  \sin \rho _{23}}{C_{23} S_{23}}
+s^e_{12}  \frac{ C_{13} S_{23} }{S_{13}  } \sin (\delta _{\nu }- \rho _{12} ) 
+s^e_{13}  \frac{ C_{13} C_{23} }{S_{13} } \sin (\delta _{\nu } - \rho _{13} ) 
\label{subsequent} \\
& \simeq \delta _{\nu } + 2 s^e_{23}  \sin \rho _{23} 
+ 5 s^e_{12} \sin (\delta _{\nu } - \rho _{12} )  
+ 5 s^e_{13} \sin (\delta _{\nu } - \rho _{13} )   \, . 
\end{align}
Here, $\r_{ij}$ denotes CP phases associated with charged-lepton mixing $s_{ij}^{e}$, and $\delta_{\nu}$ is the CP phase originating from the neutrino sector, which will be explicitly defined below. The observed mixing angles $C_{ij}$ and $S_{ij}$ are expressed in terms of the underlying lepton mixings.
For a small value of $s_{13}^{e} \lesssim 0.01$, its contribution to the Dirac phase is at most $0.05 \sim 3^{\circ}$, which is negligible compared to the expected precision $O(10^{\circ})$ of next-generation experiments. 
In this work, instead of discarding $s_{13}^{e}$, we perform an exact analysis beyond the perturbative regime. 

\end{description}

By merging the relative phases and redefining the determinant of the 2-3 mixing to be unity by the freedom of an overall phase, the MNS matrix $U_{\rm MNS}$ is rewritten as
\begin{align}
U_{\rm MNS} &\equiv \Phi_{R}^{e \dg}
U_{e}^{0 \dagger}
 \Diag{e^{i \r}}{e^{i \s}} {e^{-i\s}}
U_{\n}^{0} \Phi_{R}^{\n},  ~~~ 
U_{e}^{0 \dagger} \equiv
\begin{pmatrix}
c_{e} & - s_{e} & 0 \\
s_{e} & c_{e} & 0 \\
 0 & 0 & 1 \\
\end{pmatrix} 
\begin{pmatrix}
 1 & 0 & 0 \\
 0 & c_{\t} & -  s_{\t} \\
 0 & s_{\t} & c_{\t} \\
\end{pmatrix} \, , \nn \\
U_{\n}^{0} &\equiv
\begin{pmatrix}
 1 & 0 & 0 \\
 0 & c_{\n} & s_{\n} \\
 0 & - s_{\n} & c_{\n} \\
\end{pmatrix}
\begin{pmatrix}
c_{13} & 0 & s_{13} e^{-i \d_{\n}} \\
 0 & 1 & 0 \\
- s_{13} e^{i \d_{\n}} & 0 & c_{13} \\
\end{pmatrix}
\begin{pmatrix}
c_{12} & s_{12} & 0 \\
 - s_{12} & c_{12} & 0 \\
 0 & 0 & 1 \\
\end{pmatrix}  . 
\label{UVn}
\end{align}
Here $s_{f} \equiv \sin f\, , \, c_{f} \equiv \cos f \, , \, s_{ij} \equiv \sin \th_{ij} \, , \, c_{ij} \equiv \cos \th_{ij}$, and $\r, \s$ are arbitrary phases. 
The $\m-\t$ reflection symmetry constrains $\n = \pi/4$ and $\d_{\n} = \pm \pi/2$.

Combining the two 2-3 mixings defines new phases $\chi$ and $\h$; 
\begin{align}
& 
\begin{pmatrix}
 c_{\t} & - s_{\t} \\
 s_{\t} & c_{\t} \\
\end{pmatrix} 
\diag{e^{i \s}}{e^{ - i \s}}
\begin{pmatrix}
 c_{\n} & s_{\n} \\
 - s_{\n} & c_{\n} \\
\end{pmatrix}
= 
\begin{pmatrix}
c_{\s} c_{\n - \t} + i s_{\s} c_{\n + \t} &  c_{\s} s_{\n - \t} + i  s_{\s} s_{\n + \t}  \\
 - c_{\s} s_{\n - \t} + i s_{\s} s_{\n + \t} & c_{\s} c_{\n - \t} - i s_{\s} c_{\n + \t}  \\
\end{pmatrix}
\equiv 
\begin{pmatrix}
e^{ i \chi} c_{23} & e^{i \h} s_{23} \\
 - e^{- i \h} s_{23} & e^{- i \chi} c_{23} \\
\end{pmatrix} . 
\label{7}
\end{align}
Relations between these phases $\chi, \h$ and parameters $\s, \n, \t$ are
\begin{align}
c_{23} c_{\chi} = c_{\s} c_{\n - \t} \, , ~~~ c_{23} s_{\chi} = s_{\s} c_{\n + \t}  \, , \label{81} \\
s_{23} c_{\h} =  c_{\s} s_{\n - \t}  \, , ~~~ s_{23} s_{\h} = s_{\s} s_{\n + \t} \, .  \label{82}
\end{align}
As we will see in later analysis, $s_{23}$ defined here is almost equal to the observed 2-3 mixing. 
Substituting $\n = \pi/4$ and expressing $s_{23}$ in terms of $\s, \t$, we obtain 
\begin{align}
s_{23}^{2} = | c_{\s} s_{\n - \t} + i  s_{\s} s_{\n + \t} |^{2} = 
{ 1 - \cos 2 \sigma \sin 2 \tau \over 2} \, , 
~~~ 
\D s_{23}^{2} \equiv {- \cos 2 \sigma \sin 2 \tau \over 2} \, . 
\label{9}
\end{align}
Here $\D s_{23}^{2}$ represents the deviation from 1/2 of $s_{23}^{2}$.
In the current global fit \cite{Esteban:2024eli}, this deviation is about $\pm 0.1$ in the $3 \s$ range with or without Super-Kamiokande (SK).  Thus, the approximation $|s_{\t}| \lesssim 0.1$ appears to be  reasonable.

By multipliying diag $(1 , 1, e^{i (\chi + \h)})$ from the left and diag $(1,1,e^{ i (\chi - \h)})$ from the right of $U_{\rm MNS}$ and removing an overall phase $e ^{ i \chi}$, $U_{\rm MNS}$ is represented as
\begin{align}
U_{\rm MNS} & =
\begin{pmatrix}
c_{e} & - s_{e} & 0 \\
s_{e} & c_{e} & 0 \\
 0 & 0 & 1 \\
\end{pmatrix} 
\begin{pmatrix}
 e^{i (\r - \chi)} & 0 & 0 \\
0 & c_{23} & s_{23} \\
0 & - s_{23} & c_{23} \\
\end{pmatrix}
\begin{pmatrix}
c_{13} & 0 & s_{13} e^{-i (\d_{\n} + \h - \chi)} \\
 0 & 1 & 0 \\
- s_{13} e^{i (\d_{\n} + \h - \chi)} & 0 & c_{13} \\
\end{pmatrix}
\begin{pmatrix}
c_{12} & s_{12} & 0 \\
 - s_{12} & c_{12} & 0 \\
 0 & 0 & 1 \\
\end{pmatrix} \, . 
\label{UVn2}
\end{align}
Therefore, the following two phases $\a$ and $\b$ are CP-violating parameters; 
\begin{align}
\a \equiv \rho - \chi \, , ~~~ \b = \d_{\n} + \h - \chi \, . 
\end{align}
From Eqs.~(\ref{81}) and (\ref{82}),
$\chi \sim \eta \sim \s$ hold for sufficiently small $\t$ and it leads to  $\a \sim \rho - \s \, , \, \b \sim \d_{\n}$. 

To obtain the exact expression for $\h-\c$, we take the imaginary part of the product of the 1-2 element and the 2-2  element in Eq.~(\ref{7}), 
\begin{align}
s_{\h - \c} = \Im  e^{i(\h - \c)} = { c_{\s} s_{\s}  \over s_{23} c_{23}} s_{2 \t} \, . 
\label{13}
\end{align}
This expression implicitly depends on $s_{\n}$ through $s_{23}$ and $c_{23}$, because $\n = \pi/4$ is not substituted. 
In the range where $\sigma$ varies from $0$ to $2\pi$, 
the correction $\h-\c$ to $\b$ is expected to be about $\pm 0.2$ at most. 
A relation of $s_{\t}$ and  the above $\D s_{23}^{2}$ is derived 
by eliminating $\s$ via Eq.~(\ref{9}). 
\begin{align}
\h - \c \simeq  \pm 2 s_{\t} \sqrt{1-\frac{ (\Delta  s_{23}^2 )^2}{s_{\t}^{2}}} \, . 
 \label{h-c}
\end{align}
This sign degree of freedom $\pm$ is inherited from that of $\sin 2 \s$.

The mixing matrix~(\ref{UVn2}) has four mixing angles $s_{e}, s_{ij}$ and two phases $\a, \b$.
Since the physical parameters consist of the three mixing angles and the Dirac phase, it is in principle possible to eliminate two additional parameters by further redefinitions. 
In the subsequent work \cite{Yang:2025hex},  
a complete field redefinition produces a formula for the Dirac phase $\d = \arg [ U_{e1} U_{e2} U_{\m 3} U_{\t 3} / U_{e3} \det U_{\rm MNS}]$. 
In this paper, instead of performing further redefinitions, we show that the Dirac phase $\delta$ is extracted by applying the addition theorem to the Jarlskog invariant. 

We will outline constraints from the three observed mixing angles. 
Detailed methods are discussed in the previous letter \cite{Yang:2024ulq}. 
First, let the absolute values of the matrix elements~(\ref{UVn2}) be equal to those of the PDG representation of the MNS matrix $|U_{\rm MNS}^{\rm PDG}| = |U_{\rm MNS}|$.
The three mixing angles $s_{ij}$ of $U_{\rm MNS}$ are fixed by choosing three conditions, 
\begin{align}
|(U_{\rm MNS}^{\rm PDG})_{\t 3}| = |(U_{\rm MNS})_{\t 3}| 
~~ & \To ~~  C_{23} C_{13} = c_{23} c_{13} \, ,  \label{U23} \\
|(U_{\rm MNS}^{\rm PDG})_{e 3}| = |(U_{\rm MNS})_{e 3}| 
~~ & \To ~~ S_{13} = | c_{e} s_{13} e^{i (\a - \b)} - c_{13} s_{23} s_{e}| \, , \label{U13} \\
|(U_{\rm MNS}^{\rm PDG})_{e 2}| = |(U_{\rm MNS})_{e 2}| 
~~ & \To ~~ C_{13} S_{12} = |- c_{12} c_{23} s_{e} + e^{i \a} c_{13} c_{e} s_{12} + e^{i \b} s_{12}  s_{23} s_{13} s_{e} | \, . \label{U12}
\end{align}
Here, the observed mixing angles of $U_{\rm MNS}^{\rm PDG}$  are denoted as $S_{ij}, C_{ij}$. 
By solving these equations sequentially, $s_{ij}$ are expressed as a function of $S_{ij}, C_{ij}$. 
Since the mixing angles of the PDG representation are taken in the first quadrant by phase redefinitions, it does not lose generality to choose positive solutions $s_{ij}, c_{ij} > 0$ for $U_{\n}$.
The $s_{23}$ is immediately solved as 
\begin{align}
s_{23} &= \sqrt{1 - {C_{13}^{2} C_{23}^{2} \over c_{13}^{2}}} \, .  
\end{align}
Thus, $s_{23} \simeq S_{23}$ holds and the deviation is the second order of $s_{13}$.
Eq.~(\ref{U13}) yields a quartic equation for $s_{13}$ (or a quadratic equation for $s_{13}^{2}$).
The two solutions of $s_{13}^{2}$ are
\begin{align}
s_{13}^{2} &= 
{ A \pm B
\over (c_{e}^4 + 2 c_{e}^2 s_{e}^2 \cos 2 (\a - \b) + s_{e}^4)} \, ,  \label{sol13} \\
A & = 
- c_{e}^2 s_{e}^2 (C_{13}^2 C_{23}^2 - 1) \cos 2 (\a - \b) 
-C_{13}^2 C_{23}^2 s_{e}^4 + c_{e}^2 S_{13}^2 - S_{13}^2 s_{e}^2 + s_{e}^4 \, ,  \\
B & =  2 c_{e} s_{e} \cos (\a - \b)
\sqrt { C_{13}^2 S_{13}^2 S_{23}^2 - c_{e}^2 s_{e}^2 (C_{13}^2 C_{23}^2 - 1)^2 \sin^2(\a - \b) } \, . 
\end{align}
Since the sign degrees of freedom for $s_{13}^{2}$ are absorbed into $\cos (\a - \b)$, we choose the solution of $A+B$ that reproduces Eq.~(\ref{U13}), $S_{13} = c_{e} s_{13} - c_{13} s_{23} s_{e}$ in the limit of $\a - \b \to 0$.
The same is true for complicated solutions of $s_{12}$. 

Substituting these solutions, the remaining free parameters are the 1-2 mixing $s_{e}$ of charged leptons  and the two phases $\a, \b$. 
Finally, we will organize the sign degrees of freedom except for $c_{ij}, s_{ij}$.
In Eq.~(\ref{UVn2}), by multiplying diag $(-1 \, , \, -1 \, , 1)$ from the left,  the sign of $c_{e}$ is taken to be positive because it is changed without changing the sign of $s_{ij}, c_{ij}$.
Furthermore, we can change the sign of $s_{e}$ by multiplying $(1 \, , \, -1 \, , 1)$ from the left and redefining $\a \to \a + \pi$. Therefore, it does not lose generality to set $s_{e} > 0$. Later, we can specifically check that the sign degree of freedom of the Dirac CP phase $\d$ depends only on $s_{e} c_{e} \sin \a$. 

\section{Testability by DUNE and T2HK}

We will discuss how this scenario is verified by T2HK and DUNE in the near future.
The CP violation is evaluated from the Jarlskog invariant \cite{Jarlskog:1985ht}; 
\begin{align}
J & =  {\rm Im} \,  [U_{\m 2} U_{\t 3} U_{\m 3}^{*} U_{\t 2}^{*} ] \\
& = c_{12} c_{13} c_{23}  s_{12} [c_{13} s_{13} s_{23} (c_{e}^2 - s_{e}^2) \sin \b + (c_{23}^2 s_{13}^2-c_{13}^2 s_{23}^2) s_{e} c_{e}  \sin \a ]  \\
& + c_{13} c_{23} s_{13} s_{23} s_{e} c_{e}  [ c_{23}  (c_{12}^2-s_{12}^2) \sin(\a-\b) - c_{12} s_{12}  s_{13} s_{23} \sin(\a-2 \b)] \, . 
 \label{exp}
\end{align}
Compared to the previous letter, coefficients $s_{12} c_{12}$ is added in subleading terms proportional to $s_{13}^{2} \sin \a$ and $s_{13}^{2} \sin (\a - 2 \b)$. 
Since a redefinition $\a \to \a + \pi$ yields an interchange of the sign of $s_{e} c_{e}$,
this sign degree of freedom is absorbed by $\a$. 

The three mixing angles of the normal hierarchy (NH) without SK are referred from the latest global fit \cite{Esteban:2024eli}.
\begin{align}
& S_{12}^{2} = \sin^{2} \th_{12}^{\rm NH} = 0.307 \, , ~~ S_{23}^{2} = \sin^{2} \th_{23}^{\rm NH} = 0.561 \, , ~~~ S_{13}^{2} =  \sin^{2} \th_{13}^{\rm NH} = 0.02195 \, . 
\end{align}
The reason for using NH without SK is that the values of inverted hierarchy (IH) with or without SK are close to these values.  

\begin{figure}[t]
\begin{center}
 \includegraphics[width=16cm]{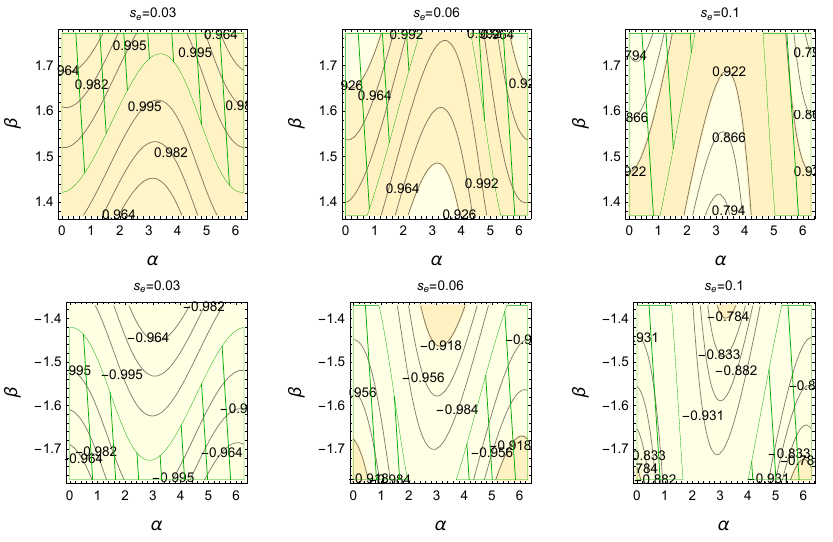}
\caption{Plots of $\sin \d$ of the MNS matrix $U_{\rm MNS}$ for a given 1-2 mixing $s_{e}$ and phases $\a, \b = \pm \pi/2 \pm 0.2$. The green-shaded regions represent $\cos \d < 0$.
}
\end{center}
\end{figure}

Fig.~1 shows plots of $\sin \d$ calculated from $J$ with $\b = \pm \pi /2 \pm 0.2$ and $s_{e} \lesssim 0.1$. 
The deviation of $\sin \d$ from unity is less than $O(s_{e})$ in most parts of regions.
This behavior is interpreted as follows. 
Substituting the three solutions of $s_{ij}$ and expanding by $s_{e}$, 
we obtain the zeroth- and first-order perturbation of $s_{e}$ for the Dirac phase $\d$ as 
\begin{align}
\sin \d^{(0)} &\simeq 
\sin \b  \, , \\
\sin \d^{(1)} &\simeq  s_{e} \cos \b \lsp
-\frac{  ( ( C_{13}^2 - 2) C_{23}^2+1) \sin  (\a -\b )}{ C_{13} S_{13} S_{23}} 
+ \frac{(1 - 2 S_{12}^2 ) C_{23}  \sin \a   }{C_{12} C_{13} S_{12} } 
 \rsp \, ,  \label{sind1} \\
 & \simeq s_{e} \cos \b \lsp
- 5 \sin  (\a -\b ) + {1\over 2} \sin \a   \rsp \, . 
\end{align}
This expression contains all first-order contributions of $s_{e}$ 
and is more accurate than the previous result \cite{Yang:2024ulq}.
Since $\cos \b$ is also small, the deviation of $\sin \d$ is small in wide regions.

On the other hand, in the region of $\sin (\a - \b) \simeq 0$, the phase $\d$ largely deviates from $\pi/2$ and the first-order contributions are insufficient to explain the behavior.
Around this region, the relative phase between two terms in Eq.~(\ref{U13}) becomes  $\a - \b \simeq \pm \pi /2$ and second-order contributions of $s_{e}$ become non-negligible, 
\begin{align}
(U_{\rm MNS})_{13} &=  c_{e} s_{13} e^{i (\a - \b)} - c_{13} s_{23} s_{e}
\simeq \pm i c_{e} s_{13}  - c_{13} s_{23} s_{e} \, , \\
S_{13}^{2} &\simeq \sqrt{(c_{e} s_{13})^{2} + (c_{13} s_{23} s_{e})^{2}} \, .  
\end{align}

To incorporate second-order effects of $s_{e}$, we reinterpret the first-order perturbation of Eq.~(\ref{sind1}) as a shift of CP phase $\e$ as follows, 
\begin{align}
\sin \e \equiv {\sin \d^{(1)} \over \cos \b} 
\simeq s_{e} \lsp - \frac{ S_{23} \sin  (\a -\b )}{ C_{13} S_{13} } 
+ \frac{(1 - 2 S_{12}^2 ) C_{23}  \sin \a   }{C_{12} C_{13} S_{12} } 
 \rsp \, . 
\end{align}
By considering Eq.~(\ref{sind1}) as the addition theorem for $\e$, 
a part of second-order perturbations of $s_{e}$ are automatically incorporated by $\cos \e$, 
\begin{align}
\sin \d \simeq \cos \e \sin \b +  \cos \b \sin \e
\simeq \sin (\b + \e)
 \, . 
\label{eqbe}
\end{align}
This expression $\sin (\b + \e)$ reproduces the Dirac phase accurately even for small $\b$.
The author also confirmed that the leading terms of second-order perturbations come from this point.

From Eq.~(\ref{eqbe}), the deviation of $\d$ from $\pm \pi/2$ is expressed by mixing angles of  charged leptons.  Together with Eq.~(\ref{h-c}),
\begin{align}
\D \d \equiv \b - \e - { \pm \pi \over 2} 
= \h - \chi + \e 
\simeq  \pm 2 s_{\t} \sqrt{1-\frac{ (\Delta  s_{23}^2 )^2}{ s_{\tau} ^2}} 
 -  s_{e} \lsp \frac{ S_{23} \sin  (\a -\b )}{ C_{13} S_{13} }  \rsp \, . 
\label{masterd}
\end{align}
Such an expansion is consistent with the general perturbative result~(\ref{subsequent}). 
In the range of $s_{e, \t} \lesssim 0.1$, 
the first term due to $\h-\chi$ is at most about 0.2, and it brings only about $\pm 10^{\circ}$ of deviation. 
On the other hand, the term involving $s_{e}$ can cause a large deviation of at most 0.5 rad $\sim 30^{\circ}$ from $S_{23}/S_{13} \sim 5$. 
In other words, the relative phase of the first generation can have a large effect on $\d$.

When the Dirac phase $\d$ is discovered or not in the near future, 
experimental interpretations of this scenario are as follows.
\begin{description}
\item[1. When the Dirac phase of about $\d = \pm \, 90^{\circ}\pm10^{\circ}$ is discovered:]

This case can be explained only by $s_{\t}$ and the relative phase $\s$ of the second and third generations.
Without accidental cancellation, a limit of $s_{e} \sin(\a - \b) \lesssim 0.2$ is imposed. 
If $\th_{23}$ is determined precisely, the magnitude of $s_{\t}$ is estimated from there.

\item[2. When the Dirac phase of about $0.8 \lesssim |\sin \d| \lesssim 0.95$ is discovered:]

This region cannot be explained by the contribution of $s_{\t}$ and $\s$ alone.
Parameters $s_{e}$ and $\cos \a$ are expected to be large and region is limited to  $\D \d \sim - 5 s_{e} \sin (\a - \b) $.

\item[3. When the Dirac phase of about $|\sin \d| \lesssim 0.8$ is not discovered:]

This boundary corresponds to $\d = \pm \, 90^{\circ}\pm40^{\circ}$. 
If the Dirac phase is not discovered in about 5 years of observations of T2HK \cite{Jesus-Valls:2024ady},  such a scenario is excluded.

\end{description}
Since the resolution around the maximum value $\d = \pm 90^{\circ}$ is about $\pm 20^{\circ}$ over the 10 years of operation of DUNE \cite{DUNE:2020jqi} and T2HK \cite{Walker:2019cxi}, 
the three situations are distinguishable to some extent. 

\subsection{CP phase of CKM matrix}

Since the observed CP phase $\d_q$ of the quarks is $O(1)$, 
it can also be interpreted as a perturbed value from the maximal CP violation, 
and it is worthwhile to perform a similar analysis for the CKM matrix. 
The best-fit values of the mixing angles and CP phase from the latest UTfit are \cite{UTfit:2022hsi}, 
\begin{align}
\sin \th_{12}^{\rm CKM} &= 0.22519 \pm 0.00083 \, ,  ~~~ \sin \th_{23}^{\rm CKM} = 0.04200 \pm 0.00047 \, ,  \\
\sin \th_{13}^{\rm CKM} &= 0.003714 \pm 0.000092 \, ,  ~~~ \d_q = 1.137 \pm 0.022 \, . 
\end{align}
For some theoretical reason, we suppose that the intrinsic CP phase of the down-type quarks $\d_{d}$ is $+ \pi/2$ which is perturbatively modified  to the observed value by up-type quarks.

Since the constraint $\n = \pi/4$ does not seem to hold for quarks, the correction $\h - \c$ to $\b$ is evaluated from Eq.~(\ref{13}) as follows.
The magnitude of the 2-3 mixing $|V_{cb}| \simeq 0.04$ of the CKM matrix is roughly close to a twice ratio of the masses $m_{f}$ of down-type quarks ${m_{s} / m_{b} } \simeq 0.02 $ \cite{Xing:2011aa}. 
If contributions from up-type quarks to the CKM matrix element are approximately kept at this ratio, it is about $ 2 m_{c} /m_{t} \simeq 0.007$. 
Although this estimation is relatively ad hoc, it is a working hypothesis aimed at gaining insight into general features of the scenario.  

From Eq.~(\ref{13}), the correction from up-type quarks $\h_{u} - \c_{u}$ to $\b$ is 
\begin{align}
\h_{u} -  \c_{u} \simeq {c_{\s} s_{\s} \over 0.04}  {2 m_{c} \over m_{t}} \in [-0.18 , + 0.18] \, ,
\end{align}
and  the parameter regions are not much different from those of leptons.
From the same argument as Eq.~(\ref{masterd}), 
corrections from relative phases of the second and third generation are insufficient to explain the observed phase $\d_q \simeq 65^{\circ}, \, \sin \d_q \simeq 0.906$. 

Fig.~2 shows similar numerical plots for quarks. The solution for $\b \sim - \pi/2$ does not exist and is omitted. 
If the 1-2 mixing $s_{u}$ of up-type quarks is small enough, $\a \simeq \pi$ is favored for the above reason.
However, as $s_{u}$ increases, the 1-3 mixing $s_{13}$ for quarks becomes smaller in Eq.~(\ref{U13}), and there is a critical value such that $s_{13} = 0$; 
\begin{align}
s_{13} = 0 ~~ \To ~~ {s_{u}^{c}}  = {S_{13} \over S_{23}} \,  .
\end{align}
This value corresponds $s_{u}^{c} \simeq 0.09$ for quarks and $s_{e}^{c} \simeq 0.2$ for leptons. 
Since the contribution of $e^{i \b}$ disappears in this neighborhood, 
it requires $|\sin \a| \sim |\sin \d_{q}|$ to explain the observed large phase $\d_q$.

\begin{figure}[t]
\begin{center}
 \includegraphics[width=16cm]{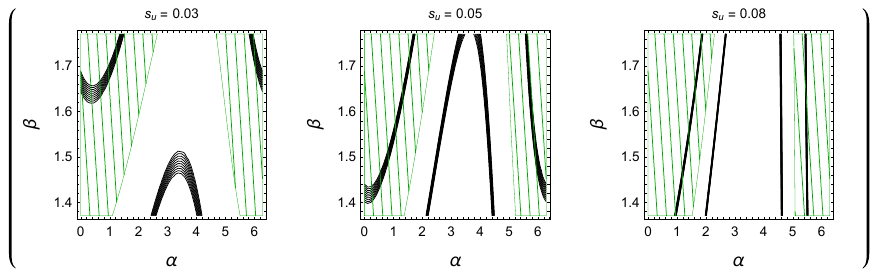}
\caption{ 
Plots of the CP violation $\sin \d_q$ for the mixing matrix of quarks $V_{\rm CKM}$. 
The green-shaded regions are excluded because they represent $\cos \d_q < 0$.
}
\end{center}
\end{figure}
%

\section{Summary}

In this paper, we generally analyze the $\mu - \tau$ reflection symmetry modified by small mixings of  charged leptons and 
how will future experiments verify deviations from the predictions of the symmetry. 
As an approximation, the left-handed diagonalization $U_{e}$ of charged leptons
 is assumed to have a similar magnitude as the CKM matrix. In other words,  the 1-3 mixing is neglected and the 1-2 and 2-3 mixing are to be about $O(0.1)$. 
 The Dirac CP phase $\delta$ of the MNS matrix is evaluated in such parameter regions. 

As a result, deviations from the predictions $\sin \theta_{23} = \pi/4$ and $\delta = \pm \pi/2$ depend on  relative CP phases between $U_{e}$ and diagonalization of neutrinos $U_{\n}$.
While phases of the second and third generations cause only about $\pm 10^{\circ}$ deviations for the Dirac phase $\d$, the phase of the first generation can cause up to $\pm 30^{\circ}$.
This flavor dependence is distinguished to some extent by the resolution $\pm 20^{\circ}$ in 10 years of operation of the next-generation experiments.
On the other hand, if $\d$ is not observed, such a scenario is excluded by about 5 years of observation. 

Similar results are obtained for the CKM matrix, and the observed CP phase in the quark sector can be interpreted  as a perturbed value from the maximal CP violation. 
Such an analysis can be useful in the study of $\mu-\tau$ reflection symmetry and flavor structures.


\begin{thebibliography}{10}

\bibitem{Fukugita:1986hr}
M.~Fukugita and T.~Yanagida,
\newblock Phys. Lett. {\bf B174}, 45 (1986).

\bibitem{T2K:2021xwb}
T2K collaboration, K.~Abe {\em et~al.},
\newblock Phys. Rev. D {\bf 103}, 112008 (2021), arXiv:2101.03779.

\bibitem{NOvA:2021nfi}
NOvA collaboration, M.~A. Acero {\em et~al.},
\newblock (2021), arXiv:2108.08219.

\bibitem{Esteban:2024eli}
I.~Esteban {\em et~al.},
\newblock JHEP {\bf 12}, 216 (2024), arXiv:2410.05380.

\bibitem{DUNE:2020jqi}
DUNE Collaboration, B.~Abi {\em et~al.},
\newblock Eur. Phys. J. C {\bf 80}, 978 (2020), arXiv:2006.16043.

\bibitem{Hyper-KamiokandeProto-:2015xww}
Hyper-Kamiokande Proto-Collaboration, K.~Abe {\em et~al.},
\newblock PTEP {\bf 2015}, 053C02 (2015), arXiv:1502.05199.

\bibitem{Walker:2019cxi}
Hyper-Kamiokande, J.~Walker,
\newblock PoS {\bf NuFact2019}, 040 (2019).

\bibitem{Ecker:1981wv}
G.~Ecker, W.~Grimus, and W.~Konetschny,
\newblock Nucl. Phys. B {\bf 191}, 465 (1981).

\bibitem{Ecker:1983hz}
G.~Ecker, W.~Grimus, and H.~Neufeld,
\newblock Nucl. Phys. B {\bf 247}, 70 (1984).

\bibitem{Gronau:1985sp}
M.~Gronau and R.~N. Mohapatra,
\newblock Phys. Lett. B {\bf 168}, 248 (1986).

\bibitem{Ecker:1987qp}
G.~Ecker, W.~Grimus, and H.~Neufeld,
\newblock J. Phys. A {\bf 20}, L807 (1987).

\bibitem{Neufeld:1987wa}
H.~Neufeld, W.~Grimus, and G.~Ecker,
\newblock Int. J. Mod. Phys. A {\bf 3}, 603 (1988).

\bibitem{Ferreira:2009wh}
P.~Ferreira, H.~E. Haber, and J.~P. Silva,
\newblock Phys. Rev. D {\bf 79}, 116004 (2009), arXiv:0902.1537.

\bibitem{Feruglio:2012cw}
F.~Feruglio, C.~Hagedorn, and R.~Ziegler,
\newblock JHEP {\bf 07}, 027 (2013), arXiv:1211.5560.

\bibitem{Holthausen:2012dk}
M.~Holthausen, M.~Lindner, and M.~A. Schmidt,
\newblock JHEP {\bf 04}, 122 (2013), arXiv:1211.6953.

\bibitem{Ding:2013bpa}
G.-J. Ding, S.~F. King, and A.~J. Stuart,
\newblock JHEP {\bf 12}, 006 (2013), arXiv:1307.4212.

\bibitem{Girardi:2013sza}
I.~Girardi, A.~Meroni, S.~Petcov, and M.~Spinrath,
\newblock JHEP {\bf 02}, 050 (2014), arXiv:1312.1966.

\bibitem{Nishi:2013jqa}
C.~Nishi,
\newblock Phys. Rev. D {\bf 88}, 033010 (2013), arXiv:1306.0877.

\bibitem{Ding:2013hpa}
G.-J. Ding, S.~F. King, C.~Luhn, and A.~J. Stuart,
\newblock JHEP {\bf 05}, 084 (2013), arXiv:1303.6180.

\bibitem{Feruglio:2013hia}
F.~Feruglio, C.~Hagedorn, and R.~Ziegler,
\newblock Eur. Phys. J. C {\bf 74}, 2753 (2014), arXiv:1303.7178.

\bibitem{Chen:2014wxa}
P.~Chen, C.-C. Li, and G.-J. Ding,
\newblock Phys. Rev. D {\bf 91}, 033003 (2015), arXiv:1412.8352.

\bibitem{Ding:2014ora}
G.-J. Ding, S.~F. King, and T.~Neder,
\newblock JHEP {\bf 12}, 007 (2014), arXiv:1409.8005.

\bibitem{Ding:2014hva}
G.-J. Ding and Y.-L. Zhou,
\newblock JHEP {\bf 06}, 023 (2014), arXiv:1404.0592.

\bibitem{Chen:2014tpa}
M.-C. Chen, M.~Fallbacher, K.~Mahanthappa, M.~Ratz, and A.~Trautner,
\newblock Nucl. Phys. B {\bf 883}, 267 (2014), arXiv:1402.0507.

\bibitem{Li:2015jxa}
C.-C. Li and G.-J. Ding,
\newblock JHEP {\bf 05}, 100 (2015), arXiv:1503.03711.

\bibitem{Turner:2015uta}
J.~Turner,
\newblock Phys. Rev. D {\bf 92}, 116007 (2015), arXiv:1507.06224.

\bibitem{Rodejohann:2017lre}
W.~Rodejohann and X.-J. Xu,
\newblock Phys. Rev. D {\bf 96}, 055039 (2017), arXiv:1705.02027.

\bibitem{Penedo:2017vtf}
J.~Penedo, S.~Petcov, and A.~Titov,
\newblock JHEP {\bf 12}, 022 (2017), arXiv:1705.00309.

\bibitem{Nath:2018fvw}
N.~Nath, R.~Srivastava, and J.~W. Valle,
\newblock Phys. Rev. D {\bf 99}, 075005 (2019), arXiv:1811.07040.

\bibitem{Yang:2021smh}
M.~J.~S. Yang,
\newblock Nucl. Phys. B {\bf 972}, 115549 (2021), arXiv:2103.12289.

\bibitem{Yang:2021xob}
M.~J.~S. Yang,
\newblock PTEP {\bf 2022}, 013B12 (2021), arXiv:2104.12063.

\bibitem{Ohki:2023zsn}
H.~Ohki and S.~Uemura,
\newblock JHEP {\bf 10}, 213 (2024), arXiv:2310.16710.

\bibitem{Harrison:2002et}
P.~F. Harrison and W.~G. Scott,
\newblock Phys. Lett. {\bf B547}, 219 (2002), arXiv:hep-ph/0210197.

\bibitem{Grimus:2003yn}
W.~Grimus and L.~Lavoura,
\newblock Phys. Lett. {\bf B579}, 113 (2004), arXiv:hep-ph/0305309.

\bibitem{Grimus:2005jk}
W.~Grimus, S.~Kaneko, L.~Lavoura, H.~Sawanaka, and M.~Tanimoto,
\newblock JHEP {\bf 01}, 110 (2006), arXiv:hep-ph/0510326.

\bibitem{Joshipura:2007sf}
A.~S. Joshipura and B.~P. Kodrani,
\newblock Phys. Lett. {\bf B670}, 369 (2009), arXiv:0706.0953.

\bibitem{Adhikary:2009kz}
B.~Adhikary, A.~Ghosal, and P.~Roy,
\newblock JHEP {\bf 10}, 040 (2009), arXiv:0908.2686.

\bibitem{Joshipura:2009tg}
A.~S. Joshipura, B.~P. Kodrani, and K.~M. Patel,
\newblock Phys. Rev. {\bf D79}, 115017 (2009), arXiv:0903.2161.

\bibitem{Xing:2010ez}
Z.-z. Xing and Y.-L. Zhou,
\newblock Phys. Lett. {\bf B693}, 584 (2010), arXiv:1008.4906.

\bibitem{Ge:2010js}
S.-F. Ge, H.-J. He, and F.-R. Yin,
\newblock JCAP {\bf 1005}, 017 (2010), arXiv:1001.0940.

\bibitem{He:2011kn}
H.-J. He and F.-R. Yin,
\newblock Phys. Rev. D {\bf 84}, 033009 (2011), arXiv:1104.2654.

\bibitem{Gupta:2011ct}
S.~Gupta, A.~S. Joshipura, and K.~M. Patel,
\newblock Phys. Rev. {\bf D85}, 031903 (2012), arXiv:1112.6113.

\bibitem{Grimus:2012hu}
W.~Grimus and L.~Lavoura,
\newblock Fortsch. Phys. {\bf 61}, 535 (2013), arXiv:1207.1678.

\bibitem{He:2012yt}
H.-J. He and X.-J. Xu,
\newblock Phys. Rev. D {\bf 86}, 111301 (2012), arXiv:1203.2908.

\bibitem{Joshipura:2015dsa}
A.~S. Joshipura and K.~M. Patel,
\newblock Phys. Lett. {\bf B749}, 159 (2015), arXiv:1507.01235.

\bibitem{Xing:2015fdg}
Z.-z. Xing and Z.-h. Zhao,
\newblock Rept. Prog. Phys. {\bf 79}, 076201 (2016), arXiv:1512.04207.

\bibitem{He:2015afa}
X.-G. He,
\newblock Chin. J. Phys. {\bf 53}, 100101 (2015), arXiv:1504.01560.

\bibitem{Chen:2015siy}
P.~Chen, G.-J. Ding, F.~Gonzalez-Canales, and J.~W.~F. Valle,
\newblock Phys. Lett. {\bf B753}, 644 (2016), arXiv:1512.01551.

\bibitem{He:2015xha}
H.-J. He, W.~Rodejohann, and X.-J. Xu,
\newblock Phys. Lett. {\bf B751}, 586 (2015), arXiv:1507.03541.

\bibitem{Samanta:2017kce}
R.~Samanta, P.~Roy, and A.~Ghosal,
\newblock JHEP {\bf 06}, 085 (2018), arXiv:1712.06555.

\bibitem{Nishi:2018vlz}
C.~C. Nishi, B.~L. S{\'a}nchez-Vega, and G.~Souza~Silva,
\newblock JHEP {\bf 09}, 042 (2018), arXiv:1806.07412.

\bibitem{Sinha:2018xof}
R.~Sinha, P.~Roy, and A.~Ghosal,
\newblock Phys. Rev. {\bf D99}, 033009 (2019), arXiv:1809.06615.

\bibitem{Huang:2018fog}
G.-Y. Huang, Z.-Z. Xing, and J.-Y. Zhu,
\newblock Universe {\bf 4}, 141 (2018).

\bibitem{Xing:2019edp}
Z.-Z. Xing and D.~Zhang,
\newblock JHEP {\bf 03}, 184 (2019), arXiv:1901.07912.

\bibitem{Pan:2019qcc}
J.~Pan, J.~Sun, and X.-G. He,
\newblock Int.\ J.\ Mod.\ Phys.\ A {\bf 34}, 1950235 (2020), arXiv:1910.06688.

\bibitem{Chakraborty:2019rjc}
K.~Chakraborty, S.~Goswami, and B.~Karmakar,
\newblock Phys. Rev. D {\bf 100}, 035017 (2019), arXiv:1904.10184.

\bibitem{Liao:2019qbb}
J.~Liao, N.~Nath, T.~Wang, and Y.-L. Zhou,
\newblock Phys. Rev. D {\bf 101}, 095036 (2020), arXiv:1911.00213.

\bibitem{Yang:2020qsa}
M.~J.~S. Yang,
\newblock Phys. Lett. B {\bf 806}, 135483 (2020), arXiv:2002.09152.

\bibitem{Zhao:2021dwc}
Z.-h. Zhao,
\newblock Eur. Phys. J. C {\bf 82}, 436 (2022), arXiv:2111.12639.

\bibitem{Bao:2022kon}
H.-C. Bao, X.-Y. Zhao, and Z.-h. Zhao,
\newblock Commun. Theor. Phys. {\bf 74}, 055201 (2022).

\bibitem{Xing:2022uax}
Z.-z. Xing,
\newblock Rept. Prog. Phys. {\bf 86}, 076201 (2023), arXiv:2210.11922.

\bibitem{Zhao:2024zgx}
Z.-h. Zhao, H.-Y. Shi, and Y.~Shao,
\newblock Phys. Rev. D {\bf 109}, 115001 (2024), arXiv:2402.14441.

\bibitem{Huang:2024xiq}
J.~Huang,
\newblock Phys. Lett. B {\bf 856}, 138898 (2024), arXiv:2405.20871.

\bibitem{Kumar:2024zfb}
R.~Kumar, N.~Nath, and R.~Srivastava,
\newblock JHEP {\bf 12}, 036 (2024), arXiv:2406.00188.

\bibitem{Xing:2017cwb}
Z.-z. Xing and J.-y. Zhu,
\newblock Chin. Phys. {\bf C41}, 123103 (2017), arXiv:1707.03676.

\bibitem{Nath:2018hjx}
N.~Nath, Z.-z. Xing, and J.~Zhang,
\newblock Eur. Phys. J. {\bf C78}, 289 (2018), arXiv:1801.09931.

\bibitem{Zhao:2017yvw}
Z.-h. Zhao,
\newblock JHEP {\bf 09}, 023 (2017), arXiv:1703.04984.

\bibitem{Liu:2017frs}
Z.-C. Liu, C.-X. Yue, and Z.-h. Zhao,
\newblock JHEP {\bf 10}, 102 (2017), arXiv:1707.05535.

\bibitem{Nath:2018xkz}
N.~Nath,
\newblock Phys. Rev. D {\bf 98}, 075015 (2018), arXiv:1805.05823.

\bibitem{Duarah:2020zjo}
C.~Duarah,
\newblock Phys. Lett. B {\bf 815}, 136119 (2021), arXiv:2006.01639.

\bibitem{Hyodo:2025xug}
Y.~Hyodo and T.~Kitabayashi,
\newblock (2025), arXiv:2502.18029.

\bibitem{Xing:2005ur}
Z.-z. Xing,
\newblock Phys. Lett. B {\bf 618}, 141 (2005), arXiv:hep-ph/0503200.

\bibitem{Farzan:2006vj}
Y.~Farzan and A.~{\relax Yu}. Smirnov,
\newblock JHEP {\bf 01}, 059 (2007), arXiv:hep-ph/0610337.

\bibitem{Hochmuth:2007wq}
K.~A. Hochmuth, S.~T. Petcov, and W.~Rodejohann,
\newblock Phys. Lett. B {\bf 654}, 177 (2007), arXiv:0706.2975.

\bibitem{Ge:2011qn}
S.-F. Ge, D.~A. Dicus, and W.~W. Repko,
\newblock Phys. Rev. Lett. {\bf 108}, 041801 (2012), arXiv:1108.0964.

\bibitem{Ge:2011ih}
S.-F. Ge, D.~A. Dicus, and W.~W. Repko,
\newblock Phys. Lett. B {\bf 702}, 220 (2011), arXiv:1104.0602.

\bibitem{Dasgupta:2014ula}
B.~Dasgupta and A.~Y. Smirnov,
\newblock Nucl. Phys. B {\bf 884}, 357 (2014), arXiv:1404.0272.

\bibitem{Petcov:2014laa}
S.~T. Petcov,
\newblock Nucl. Phys. B {\bf 892}, 400 (2015), arXiv:1405.6006.

\bibitem{Girardi:2014faa}
I.~Girardi, S.~T. Petcov, and A.~V. Titov,
\newblock Nucl. Phys. B {\bf 894}, 733 (2015), arXiv:1410.8056.

\bibitem{Girardi:2015vha}
I.~Girardi, S.~T. Petcov, and A.~V. Titov,
\newblock Eur. Phys. J. C {\bf 75}, 345 (2015), arXiv:1504.00658.

\bibitem{Delgadillo:2018tza}
L.~A. Delgadillo, L.~L. Everett, R.~Ramos, and A.~J. Stuart,
\newblock Phys. Rev. D {\bf 97}, 095001 (2018), arXiv:1801.06377.

\bibitem{Koide:2002cj}
Y.~Koide, H.~Nishiura, K.~Matsuda, T.~Kikuchi, and T.~Fukuyama,
\newblock Phys. Rev. {\bf D66}, 093006 (2002), arXiv:hep-ph/0209333.

\bibitem{Joshipura:2005vy}
A.~S. Joshipura,
\newblock Eur. Phys. J. {\bf C53}, 77 (2008), arXiv:hep-ph/0512252.

\bibitem{Yang:2020goc}
M.~J.~S. Yang,
\newblock Chin. Phys. C {\bf 45}, 043103 (2021), arXiv:2003.11701.

\bibitem{Yang:2025hex}
M.~J.~S. Yang,
\newblock Phys. Lett. B {\bf 868}, 139784 (2025), arXiv:2507.04720.

\bibitem{Yang:2024ulq}
M.~J.~S. Yang,
\newblock Phys. Lett. B {\bf 860}, 139169 (2025), arXiv:2410.08686.

\bibitem{Jarlskog:1985ht}
C.~Jarlskog,
\newblock Phys. Rev. Lett. {\bf 55}, 1039 (1985).

\bibitem{Jesus-Valls:2024ady}
Hyper-Kamiokande, C.~Jes\'us-Valls,
\newblock EPJ Web Conf. {\bf 312}, 02005 (2024).

\bibitem{UTfit:2022hsi}
UTfit, M.~Bona {\em et~al.},
\newblock Rend. Lincei Sci. Fis. Nat. {\bf 34}, 37 (2023), arXiv:2212.03894.

\bibitem{Xing:2011aa}
Z.-z. Xing, H.~Zhang, and S.~Zhou,
\newblock Phys. Rev. D {\bf 86}, 013013 (2012), arXiv:1112.3112.

\end{thebibliography}

\end{document}